\documentclass[aps,prl,twocolumn,showpacs,superscriptaddress]{revtex4}

\usepackage{graphicx}         
\usepackage{bm}               
\usepackage{amssymb}          
\usepackage{amsmath}          
\usepackage{CJK}              

\newcommand {\eV}          {\,\rm eV}

\newcommand {\kms}         {\,\rm km\,\rm s^{-1}}
\newcommand {\pc}          {\,\rm pc}
\newcommand {\kpc}         {\,\rm kpc}

\newcommand {\yr}          {\,\rm yr}
\newcommand {\Myr}         {\,\rm Myr}
\newcommand {\Gyr}         {\,\rm Gyr}
\newcommand {\Msun}        {\,\rm{M}_{\odot}}
\newcommand {\psiDM}       {\psi \rm{DM}}
\newcommand {\rhodm}       {\rho_{\rm DM}}

\newcommand {\rhostarmax}  {\rho_{\rm{*,max}}}
\newcommand {\rhosol}      {\rho_{\rm sol}}
\newcommand {\rsol}        {r_{\rm sol}}
\newcommand {\tsol}        {t_{\rm sol}}

\newcommand {\Msol}        {M_{\rm sol}}
\newcommand {\Rsolstar}    {R_{\rm cs}}
\newcommand {\rtidal}      {r_{*}}
\newcommand {\mEII}        {M_{\rm EII}}
\newcommand {\rEII}        {r_{\rm EII}}
\newcommand {\tEII}        {T_{\rm EII}}
\newcommand {\mhalf}       {M_{\rm {1/2}}}
\newcommand {\rhalf}       {r_{\rm {1/2}}}
\newcommand {\mMW}         {M_{\rm MW}}
\newcommand {\rMW}         {R_{\rm MW}}
\newcommand {\tstaradd}    {t_{0}}
\newcommand {\mvir}        {M_{\rm vir}}
\newcommand {\rvir}        {r_{\rm vir}}

\newcommand {\fref}[1]     {Fig.~\ref{#1}}

\newcommand {\eref}[1]     {Eq.~(\ref{#1})}

\newcommand {\be}          {\begin{equation}}
\newcommand {\ee}          {\end{equation}}

\begin{document}
\begin{CJK*}{UTF8}{bkai}

\title{Soliton Random Walk and the Cluster-Stripping Problem in Ultralight Dark Matter}

\author{Hsi-Yu Schive (薛熙于)}
\email{hyschive@phys.ntu.edu.tw}
\affiliation{Department of Physics, National Taiwan University, Taipei 10617, Taiwan}
\affiliation{Institute of Astrophysics, National Taiwan University, Taipei 10617, Taiwan}
\affiliation{Center for Theoretical Physics, National Taiwan University, Taipei 10617, Taiwan}

\author{Tzihong Chiueh (闕志鴻)}
\affiliation{Department of Physics, National Taiwan University, Taipei 10617, Taiwan}
\affiliation{Institute of Astrophysics, National Taiwan University, Taipei 10617, Taiwan}
\affiliation{Center for Theoretical Physics, National Taiwan University, Taipei 10617, Taiwan}

\author{Tom Broadhurst}
\affiliation{Department of Theoretical Physics, University of the Basque Country UPV/EHU, Bilbao, Spain}
\affiliation{Ikerbasque, Basque Foundation for Science, Bilbao, Spain}

\date{\today}

\begin{abstract}
Simulations of ultralight, $\sim 10^{-22}\eV$, bosonic dark matter
exhibit rich wave-like structure, including a soliton core within a
surrounding halo that continuously self-interferes on the de Broglie scale.
We show here that as an inherent consequence, the soliton undergoes a
confined random walk at the base of the halo potential.
This is significant for the fate of the ancient central star cluster
in Eridanus II, as the agitated soliton gravitationally shakes the
star cluster in and out of the soliton on a time scale of $\sim 100\Myr$,
so complete tidal disruption of the star cluster can occur within $\sim 1\Gyr$.
This destructive effect can be mitigated by tidal stripping of the halo
of Eridanus II, thereby reducing the agitation, depending on its orbit
around the Milky Way. Our simulations show the Milky Way tide affects
the halo much more than the soliton, so the star cluster in Eridanus II
can survive for over $5\Gyr$ within the soliton if it formed after significant
halo stripping.
\end{abstract}

\maketitle
\end{CJK*}

\textit{Introduction}.
A Bose-Einstein condensate of ultralight bosons with mass $m \sim 10^{-22}\eV$
has become a viable interpretation of dark matter (DM)
\cite{Widrow1993,Hu2000,Peebles2000,Goodman2000,Matos2001,Marsh2016,Schive2014a,Hui2017},
often termed fuzzy DM (FDM) or wave DM ($\psiDM$).
This has the desirable effect of suppressing dwarf galaxies \cite{Bullock2017}
as the uncertainty principle counters self-gravity below the de Broglie wavelength,
and provides the same large-scale structure as the standard cold DM
model \cite{Schive2014a,Marsh2014}.

Pioneering $\psiDM$ simulations have revealed pervasive, wave-like structure,
with a soliton core at the base of every halo, surrounded by turbulent
density fluctuations that self-interfere \cite{Schive2014a,Schive2014b}.
The soliton is a prominent flat-topped overdensity comparing favorably
with dwarf galaxies \cite{Schive2014a,Calabrese2016,Chen2017,Robles2019}
but in tension with the HI based smoothly rising rotation curves
of more massive galaxies \cite{Bernal2018,Bar2018,Robles2019}.
The ambient density fluctuations are fully modulated, which may
measurably heat stars \cite{Amorisco2018,BarOr2019,Church2019,Amr2020}.
Direct Compton scale oscillations for pulsars residing within dense solitons
may also be detectable \cite{DeMartino2017}.

The existence of a central star cluster in the ultra-faint dwarf
galaxy Eridanus II (Eri II) provides a unique probe for constraining
DM models \cite{Brandt2016}. It has been suggested that the
gravitational heating from the oscillations of soliton peak density should
destroy the star cluster completely for
$10^{-21}\eV \lesssim \mathnormal{m} \lesssim 10^{-19}\eV$ \cite{Marsh2019}.
In this Letter, using self-consistent $\psiDM$ simulations, we show
that the central soliton exhibits random-walk behavior, which can lead
to complete tidal disruption of the star cluster within $\sim 1\Gyr$ even
for $m \sim 10^{-22}\eV$. This problem is distinctly different
and potentially more serious than the gravitational heating problem above.

\begin{figure*}[ht!]
\includegraphics[width=\textwidth]{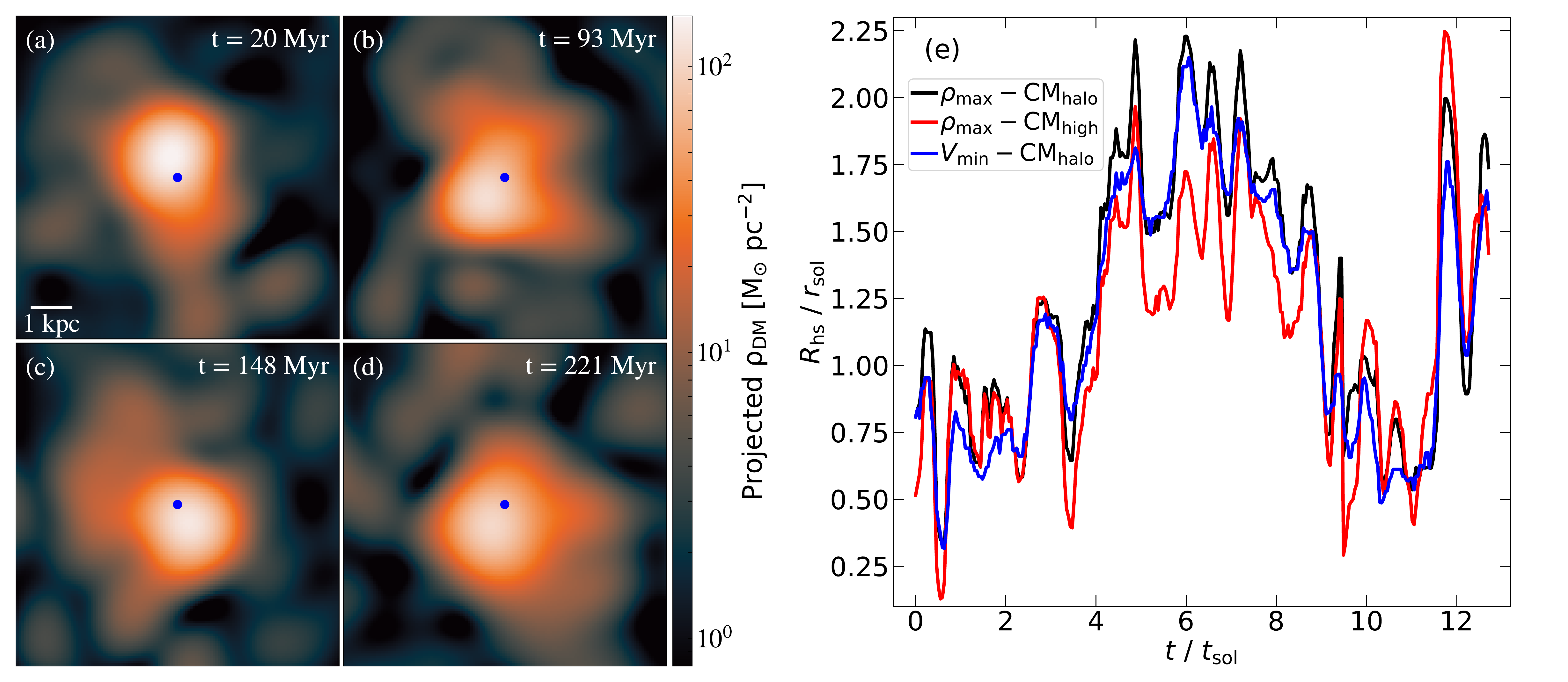}
\caption{
Soliton random walk.
(a)--(d) Projected DM density in a 4 kpc thick slab centered on
the halo center of mass (${\rm CM_{halo}}$; filled circle). Bright region (with a surface density
$\gtrsim 50\Msun\pc^{-2}$) represents the soliton. Offset between the halo
and soliton centers is manifest and constantly changing.
(e) Separation between the halo and soliton centers ($R_{\rm hs}$) as a function of time,
normalized to the characteristic length scale ($\rsol \sim 0.7\kpc$)
and time scale ($\tsol \sim 1\times10^2\Myr$) of the soliton.
Soliton random motion exhibits similar length and time scales
as $\rsol$ and $\tsol$, respectively.
These results are insensitive to the specific definitions of centers,
which we demonstrate with four different choices:
(i) ${\rm CM_{halo}}$,
(ii) CM of high-density regions with $\rhodm > 10^4\bar{\rho}_{\rm DM}$ (${\rm CM_{high}}$),
(iii) positions of the soliton peak density ($\rho_{\rm max}$), and
(iv) potential minimum ($V_{\rm min}$).
The latter two closely follow each other due to the massive compact soliton.
}
\label{fig:soliton-random-motion}
\end{figure*}

\textit{Simulation setup}.
We follow the evolution of a star cluster embedded
in the center of a $\psiDM$ halo mimicking Eri II. For the halo
component, we extract it from a cosmological simulation \cite{Schive2014b}
at redshift zero, with a virial mass of $\mvir \sim 6\times10^9\Msun$ and
a radius of $\rvir \sim 50\kpc$.
It hosts a soliton with a half-density radius of $\rsol \sim 0.7\kpc$,
an enclosed mass within $\rsol$ of $\Msol \sim 1\times10^8\Msun$, and a
peak density of $\rhosol \sim 3\times10^6 \bar{\rho}_{\rm DM} \sim 0.1\Msun\pc^{-3}$
where $\bar{\rho}_{\rm DM} \sim 4\times10^{-8} \Msun\pc^{-3}$ represents
the mean DM density. The mass within $280\pc$ is $\sim 1\times10^7\Msun$,
consistent with Eri II that has a mass of $\mhalf = 1.2_{-0.3}^{+0.4}\times10^7\Msun$
within a half-light radius of $\rhalf = 277\pm14\pc$
(corresponding to a velocity dispersion of $6.9\kms$) \cite{Li2017}.

The central star cluster in Eri II has a mass of $\mEII \sim 2\times10^3\Msun$,
a half-light radius of $\rEII \sim 13\pc$, an age of $\tEII \gtrsim 3\Gyr$,
and a very shallow density profile \cite{Crnojevic2016,Li2017}.
We model it by a Plummer sphere with a peak stellar mass density of
$\rhostarmax \sim 0.2\Msun\pc^{-3}$ and a scale radius of $\sim 20\pc$.
It corresponds to an enclosed mass within $\rEII$ of $\sim 1\times10^3\Msun$,
consistent with observations.
The star cluster is self-bound since $\rhostarmax > \rhosol$ and its
center coincides with the soliton center initially.

The governing equation of $\psiDM$ is the Schr\"{o}dinger-Poisson equation
\cite{Hu2000},
\be
\left[i\frac{\hbar}{m}\frac{\partial}{\partial t}
+ \frac{\hbar^2}{2m^2}\nabla^2-V_{\psi}\right]\psi=0,
\label{eq:Schrodinger}
\ee
\be
\nabla^2 V_{\psi}= 4\pi Gm|\psi|^2,
\label{eq:Poisson}
\ee
where $\psi$ is the wave function, $V_{\psi}$ is the gravitational potential,
$\hbar$ is the reduced Planck constant, and $G$
is the gravitational constant. $\rhodm=m|\psi|^2$ gives the
DM mass density. We adopt $m=8\times10^{-23}\eV$ throughout
this work to be consistent with \cite{Schive2014a,Schive2014b}.

To evolve $\psiDM$ and stars, we use the code \textsc{GAMER} \cite{Schive2018},
which supports adaptive mesh refinement with hybrid
parallelization. It employs an explicit second-order finite-difference method to solve
Eqs. (\ref{eq:Schrodinger}) and (\ref{eq:Poisson}) \cite{Schive2014a} and
has been extensively applied to $\psiDM$ simulations (e.g., \cite{Schive2014a,Schive2014b,Lin2018}).

We simulate a volume of size $L=250\kpc$, with a root grid $N=128^3$
and up to nine refinement levels. To resolve
the $\psiDM$ halo and soliton, cells with $\rhodm > 10^l\bar{\rho}_{\rm DM}$
are refined to level $l+1$ for $0 \le l \le 3$, leading to a resolution of
$0.12\kpc \sim (1/6)\rsol$. Furthermore, to resolve the central star cluster,
grid patches (each with $8^3$ cells) containing more than $10^3$ particles are refined, giving a maximum
resolution of $3.8\pc \lesssim (1/3)\rEII$. This high resolution
makes time-steps as small as $\Delta t \sim 150\yr$
due to the stringent Courant-Friedrichs-Lewy condition imposed
by \eref{eq:Schrodinger}. The total number of collisionless particles for sampling the
star cluster is $2\times10^5$, corresponding to a particle mass resolution
of $3.6\times10^{-2}\Msun$. We adopt isolated boundary conditions for gravity
and sponge boundary conditions for wave function.
We have validated the numerical convergence of all the presented results
by adjusting the spatial resolution by up to a factor of 4 and the particle
mass resolution by a factor of 10.
We have also confirmed that an isolated star cluster without a $\psiDM$ halo
is stable over a Hubble time in our simulations.

\textit{Isolated $\psiDM$ halo}.
We start by simulating an \emph{isolated} system by ignoring the tidal field
of the Milky Way. \fref{fig:soliton-random-motion} shows the motion
of the central soliton, revealing a confined Brownian
(random-walk-like) motion at the base of the halo potential.
This random motion exhibits a characteristic length scale similar
to the soliton radius $\rsol$ and a time scale comparable to the oscillation
period of the soliton wave function,
$\tsol \sim 120(\rhosol/0.1\Msun\pc^{-3})^{-1/2}\Myr \sim 120\Myr$.
These results are insensitive to the definitions of halo and soliton
centers (as shown in \fref{fig:soliton-random-motion}), and
the drift of the halo center of mass caused by
numerical errors is found to be of order $10^{-2}\rsol$, indicating
that this soliton random motion is not a numerical artifact.

\begin{figure*}[ht!]
\includegraphics[width=\textwidth]{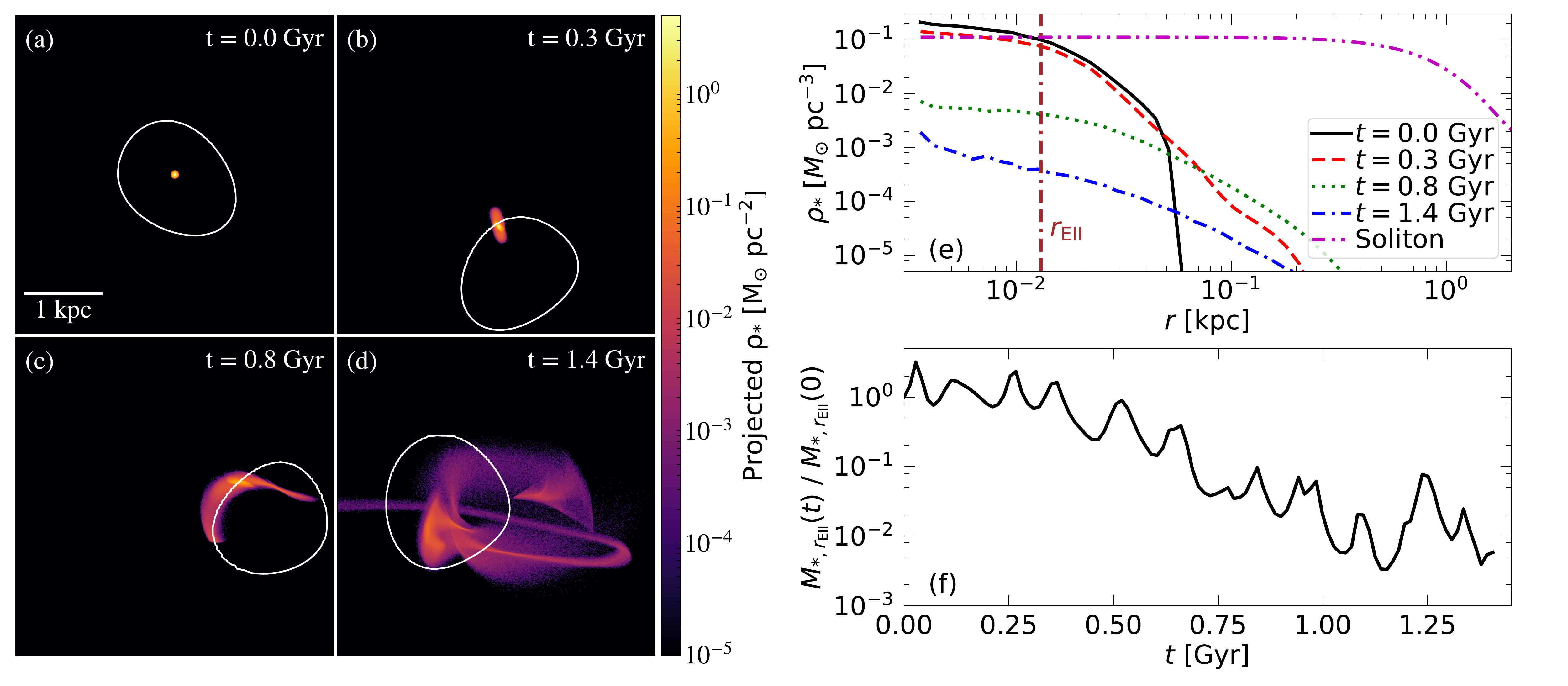}
\caption{
Tidal stripping of a star cluster caused by soliton random walk.
(a)--(d) Projected stellar mass density in a 4 kpc thick slab
centered on the initial center of the star cluster. Contours indicate
the soliton boundaries ($\sim \rsol$) where the DM density drops
by half. Although the star cluster and soliton centers coincide at the
beginning by design, the soliton random motion quickly results in a
separation between them of order $\rsol$, leading to a large
tidal field that disrupts the star cluster (see text for details).
(e) Stellar density profiles. The dash-double-dot line shows a soliton
profile for comparison. The vertical (dash-dash-dot) line indicates the
half-light radius ($\rEII$) of the central star cluster in Eri II.
(f) Enclosed stellar mass within $\rEII$ as a function of time,
normalized to its initial value. The star cluster loses $\sim 99\%$
of its original mass after $\sim 1\Gyr$.
}
\label{fig:star-tidal-stripping}
\end{figure*}

The star cluster can be treated as a tracer of the soliton as the latter is five orders of
magnitude more massive. The characteristic free-fall time of a star cluster located just outside
the soliton is $t_{\rm ff} \sim (G\rhosol)^{-1/2} \sim 50\Myr$, smaller
but comparable to $\tsol$. It suggests that the star cluster can only
loosely trace the random motion of soliton, resulting in a maximum separation
between them, denoted as $\Rsolstar$, of order $\rsol$.
\fref{fig:star-tidal-stripping}(a)--(d) illustrates this feature.

This large separation can have a great impact on the survival of the
star cluster due to tidal stripping, as shown below.
The soliton density profile features a flat core within $\sim \rsol$
and a steep outer gradient. So, for $\Rsolstar > \rsol$, a star at a
distance $\rtidal$ to the star cluster center will be subject
to a tidal field $f_{\rm tidal} \sim 2G\Msol \rtidal /\Rsolstar^3 \propto \Rsolstar^{-3}$,
assuming $\rtidal \ll \Rsolstar $.
For $\Rsolstar < \rsol$, the star cluster
will be compressed instead of tidally stripped. Therefore, tidal
stripping is most effective when $\Rsolstar \sim \rsol$, which is exactly
what happens here.
We can assess the significance of the tidal field at $\rtidal$ by computing
the ratio of $f_{\rm tidal}$ to the star cluster self-gravity $f_{\rm *}$,
\be
\frac{f_{\rm tidal}}{f_{\rm *}}
\simeq \frac{3}{2\pi}\frac{\Msol}{\rhostarmax \Rsolstar^3}
\simeq 2\frac{\rhosol}{\rhostarmax} \simeq \mathcal{O}(1),
\label{eq:tidal_ratio}
\ee
where $\rhosol \sim 0.1\Msun\pc^{-3}$ and $\rhostarmax \sim 0.2\Msun\pc^{-3}$
as described earlier.
The fact that this ratio is of order unity and independent of $\rtidal$
suggests that the entire star cluster is marginally stable and vulnerable
to tidal disruption after $t \gg t_{\rm ff}$.

This expectation is confirmed in \fref{fig:star-tidal-stripping}(e) and (f),
showing that the star cluster loses $\sim 90\%$ and $99\%$ of its original
mass within $\rEII$ after $\sim 0.7$ and $1\Gyr$, respectively. This disruption
time scale is noticeably shorter than $\tEII$ and can potentially present a serious
challenge for $\psiDM$, which we refer to as the \emph{cluster-stripping} problem.
Nevertheless, in the following, we shall further show how the Milky Way tides
may alleviate this problem by reducing the halo agitation of the soliton.

\textit{Tidally disrupted $\psiDM$ halo}.
Satellite galaxies of the Milky Way are subject to its tidal field.
Eri II, although currently at a Galactocentric distance of $\sim 370\kpc$,
may be on its second or third orbit around the Milky Way with an infall time
of $\sim 4-10\Gyr$ ago and a periapsis of $\sim 100-200\kpc$ \cite{Li2017}.
The question naturally arises as to whether the Milky Way tides can
affect the soliton random motion, and thus the survival of the central
star cluster in Eri II.

\begin{figure*}[ht!]
\includegraphics[width=\textwidth]{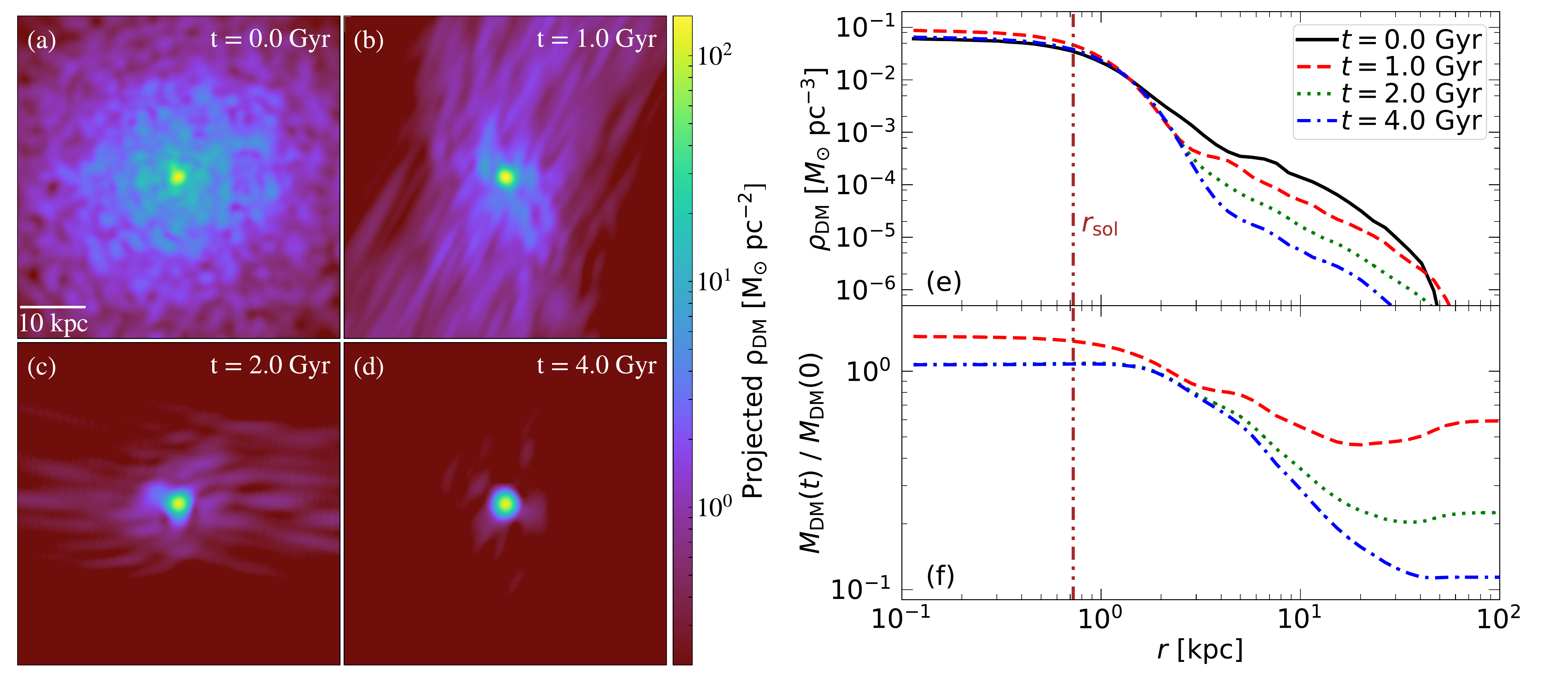}
\caption{
$\psiDM$ halo subject to the tidal field of the Milky Way.
The halo has a circular orbit of radius $100\kpc$ and a period of $\sim 3\Gyr$.
(a)--(d) Projected DM density in a 50 kpc thick slab centered on
the soliton (bright region with a surface density
$\gtrsim 50\Msun\pc^{-2}$).
(e) Halo density profiles.
(f) Halo enclosed mass profiles normalized to the initial profile.
The vertical (dash-double-dot) line indicates the soliton radius $\rsol$.
The soliton remains intact during the entire simulation period ($9\Gyr$)
while the outer region ($r \gtrsim 3\,\rsol$) is tidally disrupted after
a few Gyr.
}
\label{fig:halo-tidal-stripping}
\end{figure*}

Modeling the exact tidal stripping process of Eri II requires detailed
information about its orbital parameters, which unfortunately still suffers
from large uncertainties \cite{Fritz2018,Pace2019}.
As a proof-of-concept study, we adopt a circular orbit of radius $\rMW = 100\kpc$
for the satellite galaxy
and a point mass of $\mMW = 1\times10^{12}\Msun$ to approximate the
tidal field of the Milky Way.
To speed up the simulations, we choose a moving non-rotating coordinate
system, with the Milky Way center orbiting the coordinate origin that
coincides with the center of mass of the satellite halo,
in order to get rid of the small wavelength in $\psi$ associated with
the high orbital velocity. Assuming $r \ll \rMW$, where ${\bf r}$ and
${\bf \rMW}(t)$ are the position vectors of a simulation cell and the
Milky Way center, respectively, the tidal potential can be approximated as
\be
V_{\rm tidal}({\bf r},t) \simeq \frac{G\mMW}{2\rMW^3}
\left[ r^2 - 3\left(\frac{{\bf r}\cdot{\bf \rMW(t)}}{\rMW}\right)^2 \right].
\label{eq:tidal_MW}
\ee
The total potential is given by $V_{\rm tot}=V_{\psi} + V_{\rm tidal}$.
We evolve the system for $9\Gyr$.

\fref{fig:halo-tidal-stripping} shows the tidal stripping process of a satellite halo.
The halo surrounding the central soliton is found to be vulnerable
to tidal disruption; the density at $r \gtrsim 3\,\rsol$ decreases by more than
an order of magnitude after $\sim 2\Gyr$. In comparison, the soliton, which
is strongly gravitationally bound, is resilient to tidal disruption and stays
intact during the entire simulation period, which ensures consistency
with the observational constraint $\mhalf \sim 1\times10^7\Msun$.

\begin{figure}[ht!]
\includegraphics[width=\columnwidth]{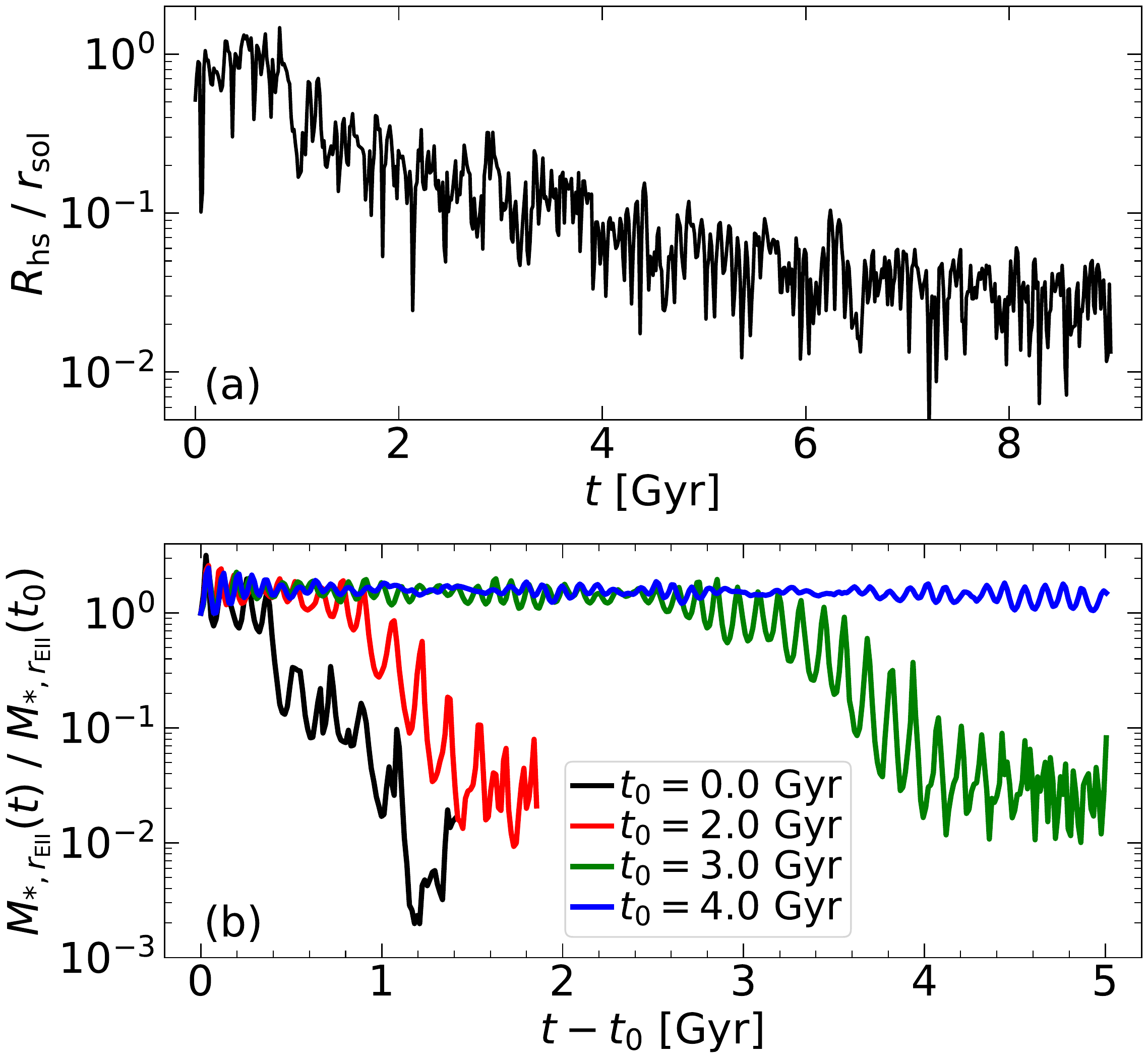}
\caption{
Soliton random walk and the resulting tidal stripping of a central
star cluster under the influence of the \emph{Milky Way potential}.
(a) Separation between the halo and soliton centers due to soliton random
walk, similar to \fref{fig:soliton-random-motion}(e).
The amplitude of random motion declines significantly as the halo being
tidally disrupted by the Milky Way tides (see \fref{fig:halo-tidal-stripping}).
(b) Enclosed stellar mass within $\rEII$ normalized to its initial value,
similar to \fref{fig:star-tidal-stripping}(f).
We add a star cluster at four different epochs ($\tstaradd$).
For each case, the horizontal axis is shifted
by $\tstaradd$ for better comparison. Clearly, the later the
star cluster is added, the more stable it is, because the random
motion becomes smaller over time and so does the induced tidal field.
For $\tstaradd = 4\Gyr$, the star cluster can remain intact for more than
$5\Gyr$, longer than the estimated minimum age of the central star cluster
in Eri II ($\sim 3\Gyr$), thus evading the cluster-stripping
problem.
}
\label{fig:star-tidal-stripping_MW-tides}
\end{figure}

During the tidal stripping of the halo, the soliton
random motion is significantly diminished.
\fref{fig:star-tidal-stripping_MW-tides}(a) shows the separation
between the halo and soliton centers as a function of time, where
we define the halo center as the center of
mass of high-density regions ($\rhodm > 10^4\bar{\rho}_{\rm DM}$)
to exclude the stripped material.
The separation, normalized to $\rsol$, reduces from $\mathcal{O}(1)$ to
$\mathcal{O}(10^{-2})$ after $\sim 5\Gyr$.
This result is not surprising given that the soliton becomes
increasingly dominant as the halo is being stripped.

In \fref{fig:star-tidal-stripping_MW-tides}(b), we assess how the above
findings affect the cluster-stripping problem by conducting four simulations
and in each of them, we add the central star cluster at a different epoch,
$\tstaradd = 0, 2, 3, 4\Gyr$. We find that the later
the star cluster is added, the more stable it is, because the tidal field
induced by soliton random motion becomes weaker over time.
For $\tstaradd=0$, the result is almost identical to the case without
considering the Milky Way tides (see \fref{fig:star-tidal-stripping})
since the tidal disruption time scale of the star cluster ($\lesssim 1\Gyr$)
is shorter than that of the halo ($\sim 2-4\Gyr$). In comparison,
for $\tstaradd=4\Gyr$, at which the ambient medium of soliton has been
largely stripped away (see \fref{fig:halo-tidal-stripping}), the star cluster
can stay intact throughout the remaining simulation time in $5\Gyr$.
This period is longer than $\tEII$ and thus free of the cluster-stripping
problem.

\textit{Concluding remarks}.
We have reported a cluster-stripping problem specific to dwarf galaxy-sized
$\psiDM$ halos hosting a central star cluster, such as Eri II. Our
unprecedentedly high-resolution simulations reveal, for the first time,
random-walk-like behavior in the soliton core. This
can displace the star cluster slightly outside the soliton radius ($\rsol$),
leading to efficient disruption of the star cluster due to the soliton tidal field.

As a possible solution, we have further demonstrated that if Eri II is
bound to the Milky Way, the tidal field of the Milky Way may readily disrupt the
outer part of a $\psiDM$ subhalo, leaving the relatively dense soliton intact.
In this case, the amplitude of soliton random motion declines
significantly, and so a star cluster forming centrally after substantial
halo removal can survive much longer. More accurate proper
motion measurements of Eri II in the future will help clarify this possibility.

We emphasize that the $\psiDM$ halo adopted here, which is consistent
with the average DM density of Eri II,
is directly extracted from a cosmological simulation with
$m \sim 1\times10^{-22}\eV$ \cite{Schive2014b}. Therefore, it appears to undermine
the claim that Eri II cannot form for $m \lesssim 8\times10^{-22}\eV$
based on the subhalo mass function \cite{Marsh2019}, which likely underestimates the total halo mass
by one to two orders of magnitude compared to our simulations.
Our study also confirms that the star cluster heating due to the oscillations of soliton peak density
is irrelevant for $m \sim 1\times10^{-22}\eV$ \cite{Marsh2019}.

The soliton peak density scales as $\rhosol \propto m^{-2}\rsol^{-4}$ \cite{Schive2014b}
and we adopt $\rsol>\rhalf$, $\rhosol \sim \rhostarmax$, and $m=8\times10^{-23}\eV$.
For $m \lesssim 5\times10^{-22}\eV$, a soliton of a fixed $\rhosol$
can still satisfy the constraint $\mhalf \sim 1\times10^7\Msun$ since $\rsol \gtrsim \rhalf$.
So the tidal field remains roughly the same (\eref{eq:tidal_ratio}).
For even larger $m$, $\rsol < \rhalf$ with a fixed $\rhosol$ and thus $\rhosol$ must increase, resulting in more efficient
tidal stripping assuming the amplitude of random motion remains $\sim \rsol$.

To test $\psiDM$ using solitons, it would be important
to consider its random motion reported here. Examples
include the dynamical friction and core stalling in dwarf galaxies
\cite{Lora2012,Hui2017,Lancaster2020}, the galactic rotation curves
\cite{Bernal2018,Bar2018,Robles2019}, and the excess mass in the
Milky Way center \cite{DeMartino2020,Bar2019,Li2020}. But it remains to be investigated how
soliton random walk changes with halo mass.

\begin{acknowledgments}
\textit{Acknowledgements}.
We thank Frank van den Bosch and Zhi Li for insightful discussions.
This research is partially supported by the Ministry of Science and Technology (MOST)
of Taiwan under the grant No. MOST 107-2119-M-002-036-MY3 and MOST 108-2112-M-002-023-MY3,
and the NTU Core Consortium project under the grant No. NTU-CC-108L893401
and NTU-CC-108L893402.
H.S. acknowledges the funding support from the Jade Mountain Young Scholar Award
No. NTU-108V0201, sponsored by the Ministry of Education, Taiwan.
\end{acknowledgments}

\bibliographystyle{h-physrev}

\newcommand {\apjl}     {Astrophys. J. Lett.}
\newcommand {\mnras}    {Mon. Not. R. Astron. Soc.}
\newcommand {\aap}      {Astron. Astrophys.}
\newcommand {\jcap}     {J. Cosmol. Astropart. Phys.}
\newcommand {\na}       {New Astron.}
\newcommand {\araa}     {Annu. Rev. Astron. Astrophys}
\newcommand {\physrep}  {Phys. Rep.}


\end{document}